\documentclass{emulateapj}

\newcommand{\Nh}{N$_{H}$}
\newcommand{\ergscm}{erg~s$^{-1}$~cm$^{-2}$}

\journalinfo{Tentatively scheduled on The Astrophysical Journal, 717:???-???, 2010 July 10 }
\slugcomment{Received 2009 December 30; Accepted 2010 May 27}

\shorttitle{Radio loudness and the AGN feedback}
\shortauthors{LA FRANCA, MELINI  \& FIORE}

\begin{document}


\title{Tools for computing the AGN feedback: radio-loudness distribution\\ and the kinetic luminosity function}

%\title{The AGN radio mode feedback from radio observations of 
%complete \\ hard X-ray AGN samples: the L$_R$/L$_X$ distribution and\\ the radio and kinetic luminosity functions}


   \author{F. La Franca$^1$,
          G.  Melini$^1$ and
          F.  Fiore$^2$ \\ ~\\
      $^1$   Dipartimento di Fisica, Universit\`a Roma Tre, 
       via della Vasca Navale 84, I-00146 Roma, Italy\\
       email: {\tt lafranca@fis.uniroma3.it, melini@fis.uniroma3.it}\\
       $^2$ INAF-Osservatorio Astronomico di Roma, 
       via Frascati 33, I-00040, Monteporzio Catone, Italy
         }


%\altaffiltext{1}{ Dipartimento di Fisica, Universit\`a Roma Tre, 
%      via della Vasca Navale 84, I-00146 Roma, Italy;
%email: {\tt lafranca@fis.uniroma3.it, melini@neve.fis.uniroma3.it}}
%\altaffiltext{2}{INAF-Osservatorio Astronomico di Roma, 
%       via Frascati 33, I-00040, Monteporzio Catone, Italy;
%       email:{\tt fiore@oa-roma.inaf.it}}

%\altaffiltext{14}{Institution, Address 00, 00000 Town, Country }


\begin{abstract}

We studied  the Active Galactic Nuclei (AGN) radio emission from a compilation of hard  X-ray selected samples, all observed
in the 1.4 GHz band. A total of more than 1600 AGN with 2-10 keV de-absorbed luminosities higher than 10$^{42}$ \ergscm\  were used.

For a sub-sample of about 50 $z$$\lesssim$0.1 AGN it was possible to reach a  $\sim$80\% fraction of radio detections and therefore, 
for the first time, it was possible to almost  completely measure  the  probability distribution function of the ratio between the radio and the X-ray luminosity
R$_X$= log(L$_{1.4}$/L$_X$), where L$_{1.4}$/L$_X$= $\nu$L$_\nu$(1.4 GHz)/L$_X$(2-10 keV).
The probability distribution function of  R$_X$ was functionally fitted as dependent
on the X-ray luminosity and redshift, $P(R_X| L_X, z)$.
It roughly spans over 6 decades  ($-7$$<$R$_X$$<$$-1$), and  does not show any sign of bi-modality.
It resulted that the probability of finding large values of the R$_X$ ratio increases with decreasing X-ray luminosities and (possibly) with
increasing redshift. 
No statistical significant difference was found between the radio properties of the X-ray absorbed (N$_H$$>$$10^{22}$ cm$^{-2}$) and unabsorbed AGN.

The measure of the probability distribution function of R$_X$ allowed us to compute the kinetic luminosity function and the kinetic energy density
which, at variance with what assumed in many galaxy evolution models,  is observed to decrease of about a factor of five at redshift below 0.5.
 About half of the kinetic energy density results to be produced by the more radio quiet (R$_X$$<$-4) AGN.

In agreement with previous estimates, the AGN efficiency $\epsilon_{kin}$ in converting the accreted mass energy into kinetic power
($L_K$$=$$\epsilon_{kin}\dot{m} c^2$) is, on average, $\epsilon_{kin}$$\simeq$$5\times 10^{-3}$. The data suggest a possible
increase of $\epsilon_{kin}$ at low redshifts.

 \end{abstract}



%%%%%%%%%%%%%%%%%%%%%%%%%%%%%%%%%%%%%%%%%%%%%%%
\section{Introduction}

One of the main questions in galaxy formation evolution studies is the
role of AGN feedback. According to popular AGN/galaxy co-evolutionary
scenarios, once central super massive black holes (SMBHs) reaches
masses $>$$10^{7}-10^8$ M$_\odot$, the AGN can heat efficiently the galaxy
interstellar matter (ISM) through winds, shocks, and high-energy
radiation (Silk \& Rees 1998, Fabian 1999), inhibiting further
accretion and star-formation and making the galaxy colors redder (see
e.g.  Cattaneo et al. 2009 for a review).

Unfortunately there are still relatively few direct observations of
AGN feedback, and its inclusion in galaxy evolution models is often
performed using adjustable parameters to obtain the observed galaxy
colors.  Indeed, the results of the hydrodynamic N-body simulations
(see, e.g. Di Matteo, Springel \& Hernquist 2005; Springel 2005;
Hopkins et al. 2005, 2006) and of the semi-analytical models of galaxy
formation and evolution (SAMs, Monaco, Salucci \& Danese 2000;
Kauffmann \& Haenhelt 2000; Volonteri, Haardt \& Madau 2003; Granato
et al.  2004; Menci et al. 2006, 2008; Croton et al. 2006; Bower et
al. 2006; Marulli et al. 2008) depend on the AGN triggering mechanism
and on the AGN feedback description.  Is the feedback at work mainly
during luminous AGN phases (the so called ``quasar'' or ``radiative''
mode, see e.g. Menci et al. 2008)? Or rather AGN feedback proceeds
continuously during the Cosmic time at a low rate (the so called
``radio'' or ``kinetic'' mode, see e.g. Croton et al. 2006, Marulli et
al. 2008)? 

In the first case the AGN feedback is associated to the the main radiative, and then X-ray, activity of
the AGN which is thought to occur during episodes of accretion of cold gas
coming from the galaxy ISM following galaxy encounters (e.g.  Barnes
\& Hernquist 1992, Cavaliere \& Vittorini 2000, Menci et al. 2008).
In this scenario the  AGN feedback is usually associated to the
radiation and  its total efficiency must be
proportional to the AGN fraction (the AGN luminosity function vs. the
galaxy luminosity function) and to the efficiency in releasing the AGN
power in the galaxy ISM.  

In the second case the radio feedback is assumed to be related to a low, uninterrupted
and constant matter accretion rate onto the central SMBH
($\sim$10$^{-5}$ M$_\odot$/yr), coming from a quiescent inflow of gas
cooling from the halo's hot atmosphere (e.g. Monaco, Salucci \& Danese 2000; Croton et al. 2006; Bower et
al. 2006). This accretion rate is too
small to contribute significantly to the bolometric output of the
AGN.  At variance, the mechanical effects of accelerated particles (jets),
which are observed to be responsible for the large cavities on the intra-cluster medium
revealed in the  X-rays (e.g. McNamara et al. 2000), are believed to significantly perturb the ISM
into radio galaxies  (Saxton et al. 2005, Sutherland \& Bicknell 2007,
Tortora et al. 2009; Krause \& Gaibler 2009 and references therein).
In this scenario, radio mode feedback total efficiency is 
thought to be proportional to the total accreted mass and then to the SMBH mass function and to the way
the energy is channelled by the SMBH into its host galaxy and it is
released in the ISM and intergalactic medium (IGM).

In the Croton et al. (2006) and Marulli et al. (2008) models, the
radio-mode feedback is more effective in suppressing the cooling
flows in the massive galaxies at late times (low redshifts). In the
Cattaneo et al. (2006) model the cooling and star-formation are
efficiently suppressed by the AGN radio feedback for haloes above a
critical mass of $\sim$10$^{12}$ M$_\sun$ below $z$$\sim$3.  Bower et
al. (2006) assume that the AGN energy injection is determined by a
self-regulating feedback loop that starts when the luminosity exceeds
some fraction of the Eddington luminosity.  Recently, Shabala \&
Alexander (2009) have presented a galaxy formation and evolution model
where the radio feedback occurs when the AGN accretion rate falls
below a certain value and enters the advection dominated accretion
flow (ADAF) regime. In their model the radio (and then kinetic) power
is assumed to scale linearly with the accretion rate.

To support the above studies several authors have compared the
mechanical (kinetic) luminosity function (LF) of radio sources to the
bolometric AGN LFs (e.g.  Best et al. 2006; Merloni \& Heinz 2008;
Shankar et al. 2008; Kording, Jester \& Fender 2008; Cattaneo \& Best
2009; Smol\v{c}i\'{c} et al. 2009).  All these works are based on the
convolution of some empirical relation between the AGN kinetic power
and its radio luminosity (see e.g. Willott et al. 1999; Best et
al. 2006; Merloni \& Heinz 2007; Birzan et al. 2008) with the AGN
radio LF. As a consequence, these works deal with the AGN as a
population and do not allow the inclusion in the models of the kinetic
power of each single source during its evolution.

In the Croton et al. (2006) and Bower et al. (2006) SAMs it is not
implemented any direct relation between the radio activity (feedback)
and the X-ray luminosity (main episodes of mass accretion).  However,
a strong correlation between the X-ray (L$_X$) and radio
luminosity of AGN is observed (e.g. Brinkmann et al. 2000), and L$_X$
is a good proxy of the SMBH accretion rate $\dot{m}$, via the
knowledge of the X-ray bolometric correction K$_X$ and the efficiency
$\epsilon$ of conversion of mass accretion into radiation
\begin{equation}
L_X = {L_{bol}\over K_X}= {\epsilon \dot{m}{c^2}\over{(1-\epsilon) K_X}},
\end{equation}
where L$_{bol}$ is the bolometric luminosity and it is assumed
$\epsilon \simeq 0.1$ (see e.g. Marconi et al. 2004; Vasudevan \&
Fabian 2009).

The aim of this paper is therefore to estimate the AGN
kinetic power linking the AGN radio emission to the  accretion
rate related to the AGN activity (the luminous phase).
Our approach is then different from most of the above described 
models, and it can be considered, in some sort, as a quasar mode feedback (i.e. related to the luminous-accreting phases), 
but associated to the emitted radio-kinetic power.
This can provide a robust quantitative root to a different kind of radio
feedback, thus guiding its self-consistent inclusion in SAMs. In this
framework, a very useful ingredient is the measure of the probability
distribution function P(R$_X$) of the ratio, R$_X$, between the AGN
radio and hard X-ray luminosity [R$_X$=$\nu$L$_\nu$(1.4
 GHz)/L$_X$(2-10 keV)].  To this purpose we used a data-set of
more than 1600 X-ray selected AGN, observed in the radio band at 1.4
GHz, to measure P(R$_X$) as a function of both luminosity and
redshift: P(R$_X |$ L$_X$, $z$).

To estimate the AGN kinetic LF and its evolution, we first
computed the radio LF by convolving P(R$_X |$ L$_X$, $z$), with the
AGN 2-10 keV LF.  We then convolved this radio LF with some of
the relations (available from the literature) between the AGN radio
and kinetic luminosities.  As sanity tests, we first compared our
results with previous studies on the relationship between the AGN
X-ray and radio luminosities, and then we checked if previous
measures of the radio LF and counts were correctly reproduced.

We adopted a flat cosmology with H$_0$ = 70 km s$^{-1}$ Mpc$^{-1}$,
$\Omega_M$=0.30 and $\Omega_\Lambda$=0.70.
Unless otherwise stated, uncertainties are quoted at the
68\% (1$\sigma$) confidence level.

%% FIG 1
  \begin{figure}[]
 \centering
  \includegraphics[width=7.5cm, angle=0]{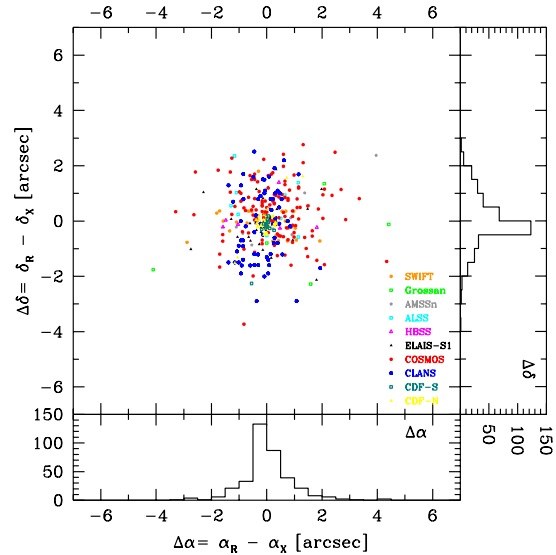}
  \caption{Distribution of the off-sets between the cross-correlated radio and X-ray sources.}
        \label{Fig_radec}
 \end{figure}

%%%%%%%%%%%%%%%%%%%%%%%%%%%%%%%%%%%%%%%%%%%%%%%
\section{The data}

As our objective was to use the R$_X$ distribution in order to estimate the kinetic (radio) luminosity of the X-ray selected AGN, we
had to measure a radio emission which was as much as possible causally linked (contemporary) with the observed X-ray activity (accretion). We then
decided to measure the radio fluxes in a region as much close as possible to the AGN, therefore minimizing the contribution 
of objects like the radio lobes in FRII sources (Fanaroff \& Riley 1974).
For this reason we built up a large data-set of X-ray selected AGN (where redshift and N$_H$ column densities estimates were available)
observed at 1.4 GHz with a $\sim$1\arcsec\ typical spatial resolution 
(in our cosmology 1\arcsec\ corresponds, at maximum, to about 8 Kpc at $z$$\sim$2).
The cross correlation of the X-ray and radio catalogues was carried out inside a region with 5\arcsec\
of radius (almost less than or equal to the size of the central part of a galaxy like ours), following a maximum likelihood algorithm as described by Sutherland and Saunders (1992) and Ciliegi et al. (2003).
In Figure \ref{Fig_radec} we show the off-sets between the X-ray and radio positions of the whole sample.
These resulted to have a root mean square (rms) of 1.4\arcsec.
Therefore we expect to have (properly) preferentially included compact FRI radio sources  in our cross-correlation and have excluded most of the contribution of the (especially bright) extended FRII radio lobes from our analysis
(see sect 3.4.2 for a discussion on the contribution of the excluded FRII sources to the radio counts).

The total AGN sample
was built up from a compilation of complete (i.e. with almost all redshift and N$_H$ measures available)
hard (mostly 2-10 keV) X-ray selected samples, with unabsorbed 2-10 keV 
luminosities higher than 10$^{42}$ erg/s\footnote{Throughout this work we  assumed that all the X-ray sources having 2-10 keV
unabsorbed luminosities higher than 10$^{42}$ erg/s are AGN (see e.g. Ranalli et al. 2003 for a study of the typical X-ray luminosities of star forming galaxies).}, as explained below.

\subsection{The bright sample. The {\it SWIFT}, {\it INTEGRAL} and {\it HEAO}-1 missions}

In order to build up a large unbiased bright AGN (low redshift) sample, we joined the AGN
samples recently generated from the {\it SWIFT} and {\it INTEGRAL} missions with the sample
published by Grossan (1992) using the  {\it HEAO}-1 data. 
In the case of sources in common, priority was given first to the
{\it SWIFT} data and then to the {\it INTEGRAL} ones and lastly to the sample of Grossan (1992).

The radio information was obtained via cross-correlation with the
1.4 GHz radio data taken from the FIRST VLA survey (Becker et al. 1995).
The FIRST images have 1.8\arcsec\ large pixels, a typical rms sensitivity of 0.15 mJy and a resolution of 5\arcsec.
In the case of no radio detection a 5$\sigma$ upper limit of 0.75 mJy was adopted.
The large area NVSS (Condon et al. 1998) and SUMSS/ATCA (Mauch et al. 2003) radio surveys have positional uncertainties significantly larger than FIRST and therefore 
were not used in our analysis.

\begin{itemize}

  \item 
{\it SWIFT.} The {\it SWIFT}  sample we used  is composed by 121 sources with high galactic latitude ($| b |$$ >$$15\degr$)
and detected with 14-195 keV fluxes brighter than 10$^{-11}$ \ergscm\
(Tueller et al. 2008). All but one of the 121 sources have a redshift and an optical spectroscopic classification available (Tueller et al. 2008). 44 out of the 121 sources have been observed by {\it FIRST}. We limited our analysis to the 40 sources with 2-10 keV unabsorbed X-ray luminosity higher than 10$^{42}$ erg/s:
there are 18 broad optical emission line type 1 and 1.5 AGN (AGN1/1.5), 13 narrow emission line type 2 AGN (AGN2), 
2 galaxies and 7 BL Lac in this sub-sample. In this and in all the other samples, all the BL Lac were excluded from our analysis as their radio
fluxes are strongly amplified by the boosting of the relativistic radio jets.
Therefore the AGN sample we used is composed by 33 sources in total (28 detected by FIRST).
\Nh\ column densities measures were provided by Tueller et al. (2008).

\item
{\it INTEGRAL.} We tried to complement the {\it SWIFT} data with the catalogue of 46 sources detected by {\it INTEGRAL} 
at a 5$\sigma$ significancy level by Beckmann et al. (2006). However, after removing two sources without N$_H$ measurements
(from Sazonov et al. 2007), and 21 sources already included in the {\it SWIFT} sample, we ended with 18 sources which were not
covered by the {\it FIRST} radio observations. Therefore, no source from the {\it INTEGRAL} catalogue from Beckmann et al. (2006) was
included in our analysis.

\item
{\it GROSSAN.} The bright sample was eventually complemented with the sample
of the {\it HEAO}-1 sources described by Grossan (1992) as revised by Brusadin (2003). 
Brusadin et al. (2003)
investigated, from the total sample of Grossan (1992), those 74 sources with 2-10 keV fluxes brighter than 2$\times$10$^{-11}$ \ergscm\ . For 52 
of these 74 sources the original optical counterparts were observed in the hard X-rays
by at least one among the {\it ASCA}, {\it Beppo-SAX} and {\it XMM-Newton} satellites. All the counterparts  
resulted to be real hard X-ray sources and a new estimate of the \Nh\ column densities were derived.
We used in this work a sub-sample of 66 objects, from the sample of Brusadin (2003), 
for which reliable measure of the \Nh\ column densities were available.

Forty-one out of these 66 sources are not included in the {\it SWIFT}
sample, and ten (10 AGN1/1.5) of them were observed by FIRST and
have 2-10 keV X-ray unabsorbed luminosities higher than 10$^{42}$ erg/s. Nine out of these last ten sources were also detected in the radio band by FIRST.

\end{itemize}

In summary, the bright sample contains 43 X-ray sources (28 AGN1, 13 AGN2, 2 galaxies; once the BL Lac were excluded)   observed in the 1.4 GHz radio band by FIRST, with 37 detections.

\subsection{AMSS}

We selected those 87 sources of the {\it ASCA} Medium Sensitivity Survey (AMSS; as in Akiyama et al. 2003) having a  S/N$>$5.5 significant detection in the X-rays, $|b|$$>$30$\degr$ and $\delta$$>$-20$\degr$; 
76 of these sources were identified as AGN: seven clusters, 3 BL Lac and one star were excluded. Forty-three of these AGN were observed by FIRST (33 AGN1, 10 AGN2), and 11 were detected.

\subsection{HBSS}

We used the 67 sources of the {\it XMM-Newton} Hard Bright Sensitivity Survey (HBSS; Della Ceca et al. 2004) with 4.5-7.5 keV fluxes brighter than 7$\times$10$^{-14}$ \ergscm\ (which corresponds to a 2-10 keV limit of  3.5$\times$10$^{-13}$ \ergscm\ if a spectral index $\alpha=0.7$, where $F_\nu\propto \nu^{-\alpha}$, is assumed).
A subsample of 62 sources were used in our analysis after the exclusion of 2 stars, one cluster and 2 sources without a  spectroscopic identification. Thirty-two sources were observed by FIRST (23 AGN1, 9 AGN2), while 6 were detected.

\subsection{ALSS}

The {\it ASCA} Large Sky Survey (ALSS) is a contiguous 7 $\deg^{2}$ strip in the North Galactic Pole region (Ueda et al. 1999);
we selected a sample with a limiting 2-10 keV flux of 1$\times$10$^{-13}$ \ergscm\  from Akiyama et al. (2000).
This sample contains 30 AGN (25 AGN1, 5 AGN2), as well as two clusters, one star and one object without spectroscopic identification, which were excluded from our analysis. All the 30 AGN were  observed by FIRST, while nine were detected.

\subsection{COSMOS}

In order to use the COSMOS survey, we used the XMM-Newton X-ray catalogue by Cappelluti et al. (2009) with a limiting 2-10 keV flux 
of $\sim$3$\times$10$^{-15}$ \ergscm\ and cross-correlated with the spectroscopic identifications by Trump et al. (2009) and the photometric
redshift estimates from Ilbert et al. (2009) and Salvato et al. (2009). The \Nh\  measures were derived by our analysis of the hardness ratios (HR). The radio data were obtained from Schinnerer et
al. (2007). In order to allow a uniform radio coverage (with 1.4 GHz rms of 15 $\mu$Jy) the central 1 deg$^2$ squared area with limits 9$^h$58\arcmin40$^s$$<$$\alpha$$<$10$^h$2\arcmin40$^s$ and 1\arcdeg42\arcmin$<$$\delta$$<$2\arcdeg42\arcmin\  was used. In this area the 
X-ray catalogue contains 712 sources.  677 are extragalactic with a redshift measure available and 2-10 keV luminosity higher than 10$^{42}$ erg/s. 389 have a spectroscopic redshift (186 AGN1, 52 AGN2, 71 emission line galaxies, ELG, 32 normal passive galaxies, GAL,  48 no class), while 288 have only a photometric redshift estimate available. We used all the radio detections
with 1.4 GHz flux limits brighter than  75 $\mu$Jy (5$\sigma$) and used the same threshold as an  upper limit for all the remaining sources even though lower flux detections were available in some cases. In total 141 out of the 677 sources, contained in our selected region of the COSMOS field, were detected in the 1.4 GHz radio band.

\begin{table}[t]
\caption{The Samples}
\begin{tabular}{lrrcccc}
\hline
\hline
\footnotesize
\smallskip
 Sample &
 N$_{1.4}$&
 N$_X$&
   $N_{1.4}\over N_X$&
F$_{1.4}$&
 F$_X$&
   $F_{1.4}\over F_X$\\

& (1) &  (2)&  (3)&  (4)& (5) & (6)\\
%\hline
\tableline

%\multicolumn{17}{c}{LDDE}\\
%\tableline
Bright       & 37 & 43 &0.86& 750 & \phantom{2.}2$\times$10$^{-11}$ & -6.3\\
AMSS       & 11 & 43 & 0.26 & 750& \phantom{2.}3$\times$10$^{-13}$& -4.5\\
ALSS        & 9 & 30 & 0.30 &750 & \phantom{2.}1$\times$10$^{-13}$ &-4.0\\
HBSS        & 6 & 32& 0.19 & 750 & 3.5$\times$10$^{-13}$ &-4.5\\
COSMOS & 141 & 677 & 0.21& \phantom{0}75&\phantom{2.}3$\times$10$^{-15}$ &-3.5\\
CLANS     & 69 & 139 & 0.50&\phantom{0}19 & \phantom{2.}3$\times$10$^{-15}$ & -4.1\\
ELAIS-S1  & 45 & 421 & 0.11& 150&\phantom{2.}2$\times$10$^{-15}$ & -3.0\\
CDF-S      & 12 & 94 & 0.13&\phantom{0}70& 2.6$\times$10$^{-16}$&-2.4\\
CDF-N      & 45 & 162 & 0.28& \phantom{0}45 & 1.4$\times$10$^{-16}$&-2.3\\
\\
Total & 375 & 1641 & 0.23 & ...  & ...& ...\\

\tableline
\end{tabular}
\tablenotetext{}{Notes. 1) Number of AGN detected at 1.4 GHz; 2) Number of X-ray AGN observed at 1.4 GHz;
3) Fraction of AGN detected at 1.4 GHz;  5) 1.4 GHz radio flux limit in $\mu$Jy units; 5) 2-10 keV flux limit in
\ergscm\ units; 6) log[$\nu$F$_\nu$(1.4 GHz)/F$_\nu$(2-10 keV)].\\}
\label{tab_samp}
\end{table}
%%%%%%%%%%%%%%%%%%%%%%%%%%%%%%%%%%%%%%%%%%%%%%%%%%%%%%%%%%%%%%%%%%%%%%%%%%%%%%%

%%FIG 2
  \begin{figure}[]
 \centering
  \includegraphics[width=6cm, angle=-90]{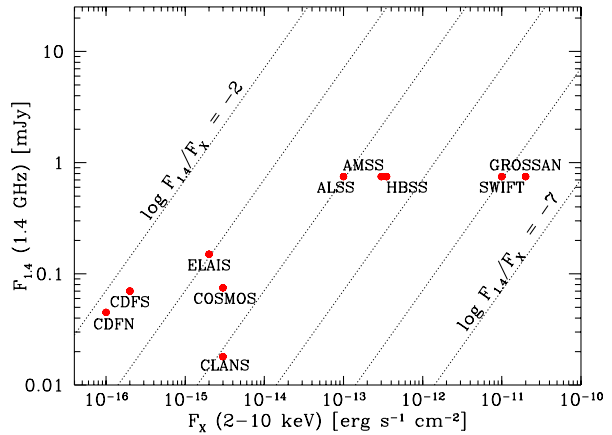}
  \caption{Lower (5$\sigma$) radio flux 1.4 GHz limits as a function of the lower X-ray 2-10 keV flux limits of the samples.
  Dashed lines show the loci with equal $\nu$F$_\nu$(1.4 GHz)/F$_\nu$(2-10 keV) ratio as shown by the labels.}
        \label{Fig_FxFr}
 \end{figure}

\subsection{ELAIS-S1}

In the European Large Area {\it ISO} Survey field S1 (ELAIS-S1) we used the catalogue of the XMM-Newton sources published by Puccetti et al. (2006), which reaches a 2-10 keV flux limit of  2$\times$10$^{-15}$ \ergscm. The spectroscopic identifications and classifications 
provided by Feruglio et al. (2008)  and Sacchi et al. (2009) were used, while the 1.4 GHz radio data (with a 5$\sigma$ limit of 150 $\mu$Jy) 
were taken from Middelberg et al. (2009).  The whole sample contains 421 extragalactic sources with a redshift measure available and 2-10 keV luminosity higher than 10$^{42}$ erg/s: 240 have been identified and classified spectroscopically (116 AGN1, 34 AGN2, 68 ELG, 22 GAL), while 181 have  a photometric redshift available.
Forty-five out of these 421 AGN were detected at 1.4 GHz above the 150 $\mu$Jy limit.

%% FIG 3
  \begin{figure}[]
 \centering
  
  \includegraphics[width=8cm, angle=0]{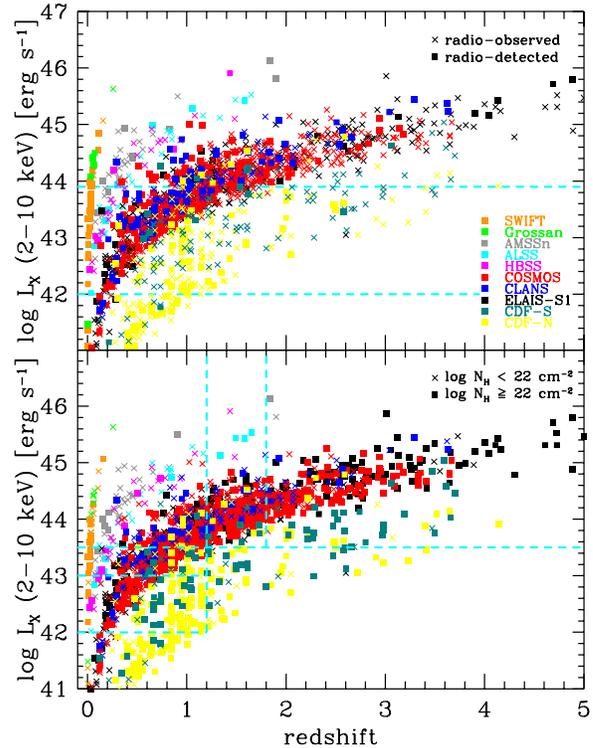}
  \caption{2-10 keV X-ray de-absorbed luminosity as a function of the redshift for all the AGN observed at 1.4 GHz
  used in this work. {\it Top}. Crosses indicate the sources observed in the radio band, while filled squares
  show those objects which have been detected in the radio. The dashed lines show the two $z$-L$_X$ regions used to measure the  distribution of R$_X$ at R$_X$$<$-4.
  {\it Bottom}.
  Crosses indicate the sources with log \Nh\ $<$ 22 cm$^{-2}$, while filled squares show those AGN
  with log \Nh\ $\geq$ 22 cm$^{-2}$. 
  The dashed lines show the five $z$-L$_X$ regions used to measure the  distribution of R$_X$ at R$_X$$\geq$-4.}
        \label{Fig_LZ}
 \end{figure}

\subsection{CLANS}

In the Chandra Lockman Area North Survey (CLANS) field the X-ray data were taken from Trouille et al. (2008), which also publish the spectroscopic and photometric
redshift measures.  Our sample consists of the sources with S/N$>$3 detections in the hard 2-8 keV band and
includes in the circular area with radius of 0.32 $\deg$ and center in: $\alpha$ 10$^h$46$\arcmin$, $\delta$ $59\arcdeg01\arcmin$ (J2000). This area contains 139
extragalactic sources with a redshift measure available and 2-10 keV luminosity higher than 10$^{42}$ erg/s: 113 sources were spectroscopically identified (58 AGN1, 34 AGN2, 16 ELG, 5 galaxies), while 29 sources have a photometric redshift available. The \Nh\ measures were derived from the analysis of the HR. Radio data were obtained from Owen et al. (2008). 
We conservatively modeled the spatial dependence of the (5$\sigma$)  radio flux limits, which vary from 18.5 $\mu$Jy in the central region up to 59.2 $\mu$Jy near the edges. Sixty-nine out of the 139 sources were radio-detected.

\subsection{CDFS}

In the {\it Chandra} Deep Field South (CDFS) the subsample of the GOODS-S X-ray sources from the catalogue of Alexander et al.
(2003) was used. The whole sample consists of the 94 point like extragalactic sources over a 2-10 keV flux
limit (at the aim point) of 2.6$\times$10$^{-16}$ \ergscm\ with a redshift measure available and a 2-10 keV luminosity higher than 10$^{42}$ erg/s. Spectroscopic identifications for 69 sources, as well as 25 photometric redshifts, were obtained from (Brusa et al. 2009b).  Twenty-nine sources were identified as AGN (17 AGN1, 12 AGN2), 39 as ELG and one as a galaxy.
The radio data were taken from Miller et al. (2008) and have a 1.4 GHz 5$\sigma$ flux limit of 70 $\mu$Jy. Twelve
sources were radio-detected.

\subsection{CDFN}

In the {\it Chandra} Deep Field North (CDFN)
we used the 296 2-8 keV X-ray sources detected by {\it Chandra} within the GOODS-N area with a 2-10 keV flux limit (at the aim point) of 
1.4$\times$10$^{-16}$ \ergscm\  by Alexander et al. (2003). To convert the 2-8 keV fluxes in the 2-10 keV band a spectral index $\alpha$$=$$0.4$ were assumed. The spectroscopic identifications were taken from Trouille et al. (2008). 
The sample consists of 162 extragalactic sources with a redshift measure available and 2-10 keV luminosity higher than 10$^{42}$ erg/s: 104 sources were spectroscopically identified (16 AGN1, 20 AGN2, 46 ELG, 22 galaxies), while 50 sources have a photometric redshift; eight other spectroscopical redshift were retrieved from literature. 
\Nh\ values were obtained from the analysis of the HR.  The radio information were obtained from
the new data reduction from Biggs \& Ivison (2006) of the 1.4 GHz VLA observation of Richards (2000).
Forty-five out of the 162 extragalactic sources were identified in the 1.4 GHz band over a (5$\sigma$) flux limit of 45 $\mu$Jy.

\subsection{The whole X-ray sample}

In summary, our radio observed hard X-ray selected extragalactic total sample contains 1641 sources
with both redshifts (either spectroscopic or photometric) and \Nh\ column densities measured, and 
with un-absorbed 2-10 keV luminosities higher than  10$^{42}$ erg/s\footnote{Other X-ray samples, 
such as HELLAS2XMM (Fiore et al. 2003; Cocchia et al. 2007), the XMM/Lockman Hole
(Brunner et al. 2008) and the XMM Medium Survey (Barcons et al. 2007),
were not included in our work 
because unbiased, homogeneous and dedicated radio 1.4 GHz observations were not available.}. 
1003 sources have \Nh\ higher than 10$^{22}$ cm$^{-2}$ (hereafter defined  ``X-ray absorbed''). 
375 (23\%) sources were detected in the 1.4 GHz radio band. See Table \ref{tab_samp} for a summary
of the main properties of all samples used. 
In Figure \ref{Fig_FxFr} we show the (5$\sigma$) radio 1.4 GHx flux limits of each survey
as a function of their deepest 2-10 keV X-ray flux limits. 
The distribution of the de-absorbed X-ray luminosity of all sources  (distinguished according to radio
detection and X-ray absorption classes) as a function of redshift is shown in Figure
\ref{Fig_LZ}. 

\section{The distribution of  the R=L$_R$/L$_X$ ratio}

\subsection{The method}

We searched for a functional fit of the probability distribution function of R$_X$, as a
function of the X-ray luminosity, L$_X$, and the redshift: $P(R_X |  L_X, z)$. The
method is based on the comparison, through $\chi^2$ estimators, of the
observed and expected numbers of AGNs (in the L$_X$-$z$-R$_X$ space) obtained by taking into
account the observational selection effects (i.e. the radio flux limits) of each sample.
Once  a  probability distribution function $P(R_X | L_X, z)$ is assumed, the number of
expected AGNs  in a given bin of the L$_X$-$z$-R$_X$ space is the
result of the sum, over the number of all AGN contained in the L$_X$-$z$ bin, of the expected
number of AGNs contained in that  bin of R$_X$, by taking into account the radio flux limits on each source.
This method reproduces the observations and consequently properly takes into account
both the radio detections and the upper limits
(see La Franca et al. 1994, 1997, 2002, 2005 for similar applications).

\subsection{The fit}

%%%%%%%%%%%%%%%%%%%%%%%%%%%%%%%%%%%%%%%%%%%%%%%%%%%%%%%%%%
%%% TAB 2
\begin{table*}[t]
\caption{Best fit parameters}
\begin{tabular}{lcccccccc}
\hline
\hline
\footnotesize
\smallskip
 Model &
 N&
 $\gamma_R$ &
 $\gamma_L$ &
 $R_C$ &
 $\alpha_L$ &
 $\alpha_z$ &
 $\chi^2/$d.o.f. &
 P$(\chi^2)$\\
%\hline
\tableline

1 - box       & $...$ & $...$& $...$& $...$& $...$& $...$& 670.17/12 & 0\\
2 - no dep & 1.0620 & 0.476 & 1.93 & -4.313 & $...$    & $...$    & \phn46.73/20 & $6.4\times 10^{-4}$\\
3 - dep z   & 1.0899 & 0.467 & 1.69 & -4.319 & $...$    & \phm{-}0.028 & \phn41.98/19 & 0.020\\
4 - dep L   & 1.0652 & 0.429 & 1.70 & -4.386 & 0.056 & $...$    & \phn30.19/19 & 0.049\\
5 - dep L,z & 1.0230 & 0.369 & 1.69 & -4.578 & 0.109 & -0.066& \phn22.77/18 & 0.200\\
\\
1$\sigma$ errors &  & $^{+0.040}_{-0.031}$ & $^{+0.18}_{-0.31}$ & $^{+0.110}_{-0.086}$ & $^{+0.019}_{-0.025}$ & $^{+0.024}_{-0.016}$ & & \\

\tableline
\\
\\
\end{tabular}
\label{tab_fit}
\end{table*}
%%%%%%%%%%%%%%%%%%%%%%%%%%%%%%%%%

\begin{figure}[]
 \centering
  \includegraphics[width=6.2cm, angle=-90]{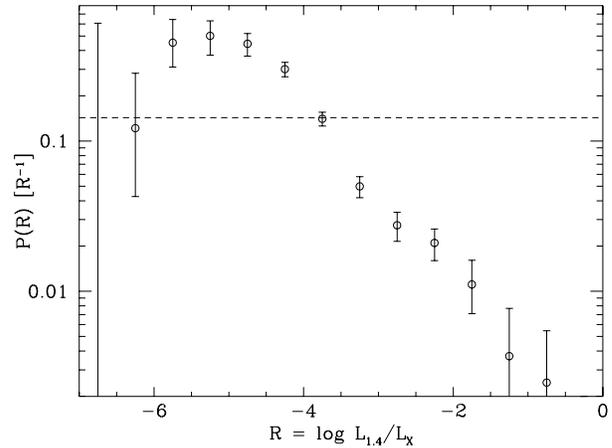}
  \caption{Probability distribution function of R$_X$ for the full sample under the assumption of a flat distribution (dashed line).}
        \label{Fig_box}
  \end{figure}

As a first test we assumed a constant (flat) probability distribution function of R$_X$ in the range -7$<$R$_X$$<$0 (see Figure \ref{Fig_box}). This distribution, although different from the true one, allows to see which is the shape of the average true distribution
via the analysis, in each bin,  of the deviations of the observed numbers of AGN from the expected ones. Figure \ref{Fig_box} shows that the average
distribution function is a-symmetrical with a long tail at large R$_X$ values.

%% FIG 4

\begin{figure}[b]
 \centering
  \includegraphics[width=8.5cm, angle=0]{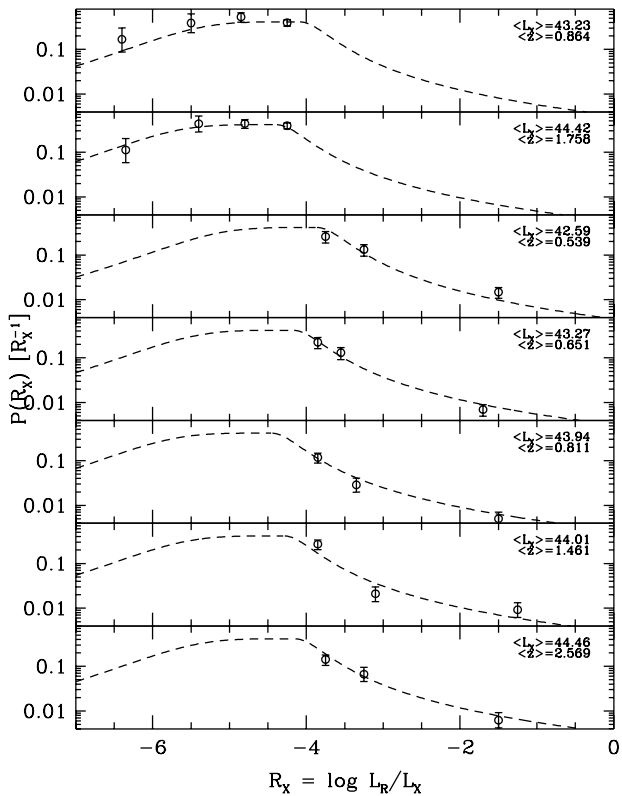}
  \caption{Probability distribution function of R$_X$ as a function of L$_X$ and $z$. The dashed line shows our best fit solution \#5 (see Table \ref{tab_fit}). The limits in the L$_X$-$z$ space of the seven  regions used are shown in Figure \ref{Fig_LZ}.}
        \label{Fig_LogDistr5}
 \end{figure}

We then divided the L$_X$-$z$-R$_X$ space into 23 independent bins, which
were chosen in order to look for possible dependences of the  $P(R_X|  L_X, z)$ shape
on the luminosity and redshift (see Figure \ref{Fig_LZ} for a representation of the L$_X$-$z$ regions) while ensuring that about, 
at least, 10 AGN were observed in each bin (see Figure \ref{Fig_LogDistr5}).

After several trials, we found  that the probability distribution function distribution of R$_X$ is, indeed, a-symmetrical, showing a maximum
at $R_X$$=$$R_0$, where the median is located.  At R$_X$ larger than $R_0$ the distribution is
fairly well represented by a Lorentz function having width $\gamma_r$, which provides a shallow decline at large R$_X$ values, while at R$_X$ smaller than $R_0$ the exponent 2 of the Lorentz function is better substituted by an exponent 4 which gives a wider shoulder at $R_X$$ \lesssim$$ R_0$ and then a steep decline at even lower R$_X$. The width of the left ($R_X$$<$$R_0$) part of the distribution is controlled by the $\gamma_l$ parameter. Therefore the probability distribution function $P(R_X|  L_X, z)$
is expressed by the following formula:
\begin{eqnarray}
P(R_X)=\left\{\begin{array}{l l}\frac{N}{A\pi\gamma_l \left[ 1 + \left(\frac{R_0(L_X,z)-R_X}{\gamma_l} \right)^4 \right]}& \mbox{($R_X<R_0$)}\\ 
\\
\frac{A\ N}{\pi\gamma_r  \left[ 1 + \left(
\frac{R_X-R_0(L_X,z)}{\gamma_r} \right)^2 \right]}
& \mbox{($R_X\geq R_0$),}\\ \end{array} \right. \
\end{eqnarray}

\noindent
where,  in order to obtain a continuous function at $R_X$$=$$R_0$ results $A = \sqrt{\gamma_r/\gamma_l}$, and
the parameter $N$ is constrained by the probability normalization requirement: $\int P(R_X| L_X, z)dR_X=1$.
For $0.3$$\leq$$z$$\leq3.0$ and $42.2$$\leq$$LogL_X$$\leq$$47.0$, we allowed to vary, as a function of $L_X$ and $z$, the position  of the maximum (median) $R_0$ of the distribution, according to the following formula:
\begin{equation}
R_0 = R_C \left[  \alpha_L (LogL_X-44)+ 1   \right] \left[  \alpha_z (z-0.5) + 1 \right].
\label{Eq_dipLZ}
\end{equation}
At redshifts and luminosities outside these ranges,  $R_0$ was kept constant,
equal to the values assumed at the limits of the ranges.

In Table \ref{tab_fit} the results of the fits carried out using this parameterization are reported.
Confidence regions of each parameter were obtained by computing $\chi^2$ at a number of values
around the best-fit solution, while leaving the other parameters free to float (see Lampton et al. 1976). The 68\% confidence regions
quoted correspond to $\Delta \chi^2$$=$$1$.

The solution without dependences on both the luminosity and the redshift is rejected by the $\chi^2$ test, while
the solutions either depending only on luminosity or redshift provide barely acceptable fits to the data.
A fairly good fit to the data (20\% $\chi^2$ probability) is instead provided by the solution \#5 where both a dependence on the luminosity and the redshift is allowed.
However, it should be noted that the parameter of the redshift dependence,
$\alpha_z$, is different from zero only at 3$\sigma$ confidence level.

\begin{figure}[]
 \centering
  \includegraphics[width=6.cm, angle=-90]{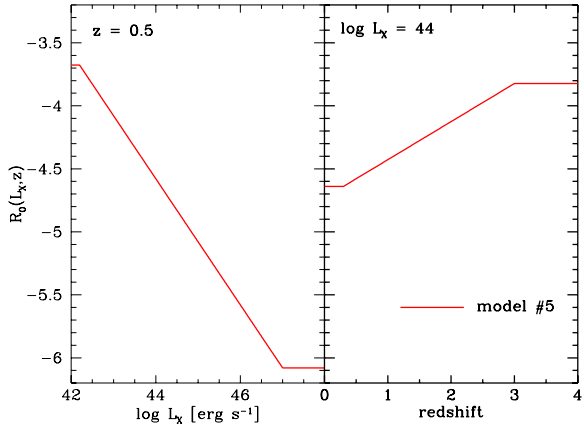}
  \caption{Dependence of the position of the median R$_0$ of the probability distribution function on L$_X$ and $z$ as computed in our best fit solution \#5 (red continuous line; see text and Table \ref{tab_fit}).}
        \label{Fig_LZ5}
 \end{figure}

\begin{figure}[b]
 \centering
  \includegraphics[width=8cm, angle=0]{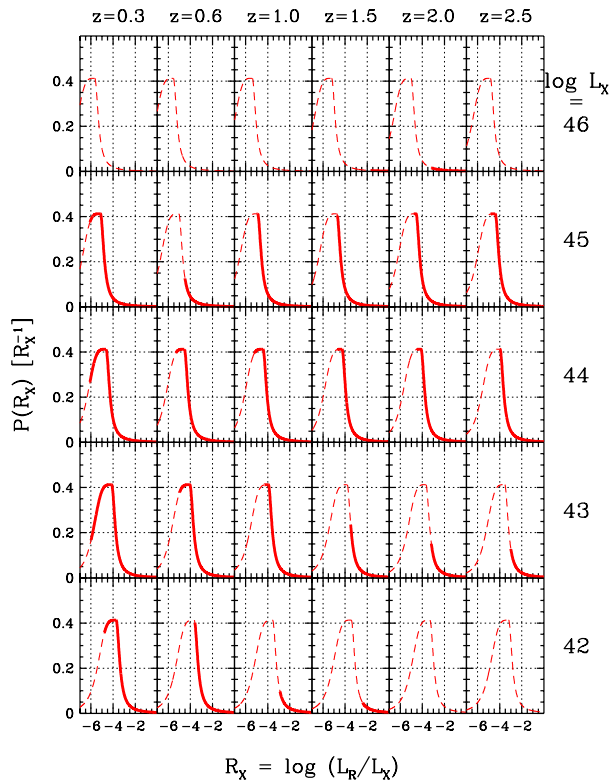}
  \caption{Probability distribution function in different bins of L$_X$ and $z$, as computed in our best fit solution \#5 (see Table \ref{tab_fit}).
    The continuous line shows the range of R$_X$ where the fit is constrained by the data, while the dashed line shows where the distribution is extrapolated.}
        \label{Fig_distr5}
 \end{figure}

The data and the shape of the best fit \#5 probability distribution function are shown in Figure \ref{Fig_LogDistr5}, while the corresponding dependences of $R_0$ on L$_X$ and $z$ are shown in Figure  \ref{Fig_LZ5}.  In Figure
\ref{Fig_distr5} we show the shape of the best fit \#5 probability distribution function in different bins of  X-ray luminosity and redshift,
with evidence (the continuous lines) to the part which is actually constrained by the data.

%% FIG 5

%%% FIG 6

%%% FIG 7

%% FIG 8
  \begin{figure}[t]
 \centering
  \includegraphics[width=8.3cm, angle=0]{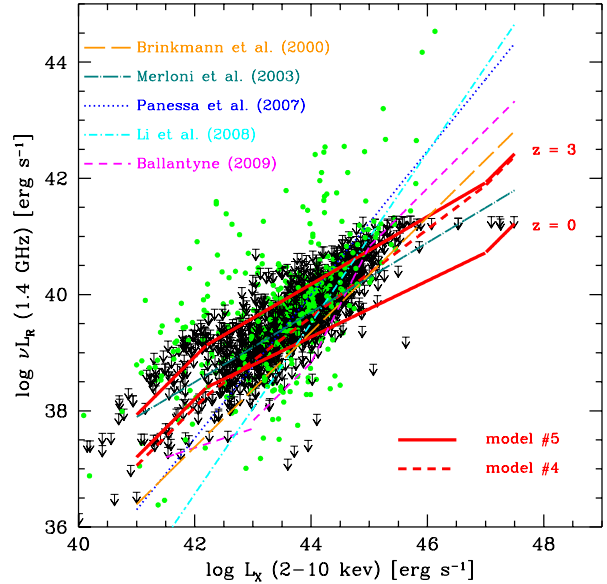}
  \caption{1.4 GHz luminosity as a function
of the intrinsic 2-10 keV luminosity of all AGN contained in our sample (only the AGN with logL$_X$$\geq$42 have been used
in our analysis).  Radio detection are shown by green dots, while radio upper limits are shown by arrows.
The dependence, according to fit \#5, of the position of the median R$_0$ of the $P(R_X | L_X, z)$ distribution (see eq. \ref{Eq_dipLZ})
at redshift 0 and 3 is shown by red continuos lines. For comparison the relations derived by Brinkmann et al. (2000; orange long dashed),
Panessa et al. (2007; blue dotted), and Ballantyne (2009; magenta dashed) are shown. The relations derived by the measure of the
AGN fundamental plane by Merloni et al. (2003) and Li et al. (2008) are shown (green dot-dashed and cyan dot-dashed lines, respectively) assuming a fixed  BH mass of 10$^8$ M$_\sun$ (see text).}
        \label{Fig_LRLX}
 \end{figure}

\subsection{Dependence on N$_H$}

We did not find any
significant dependence of the $P(R_X| L_X, z)$ distribution on the N$_H$ column densities. The sample was divided into absorbed (N$_H$$>$10$^{22}$ cm$^{-2}$) and un-absorbed AGN. For both sub-samples the $\chi^2$ test on the best fit solution \#5 (even with a different sampling in order to always observe at least 10 objects in each bin) provided probabilities larger than 10\% (up to 70\%), and was then not able to reject our  best fit distribution.
This result is in agreement with the analysis of the X-ray absorption properties of the faint radio sources of the CDFS by Tozzi et al. (2009).

\subsection{Sanity checks}

\subsubsection{The L$_R$-L$_X$ relation}

We compared our measure of the probability distribution function of R$_X$ with previous measures on the
relationship between the AGN radio and X-ray luminosities. In Figure \ref{Fig_LRLX} we show the 1.4 GHz luminosity as a function
of the intrinsic 2-10 keV luminosity for all the sources of our sample. Our analysis differs from many previous studies where single general log-log linear relations were derived. In these studies,  even when the presence of censored data, or data with errors in both axes (see e.g. La Franca et al. 1995), were
taken into account, it was assumed the presence of a symmetrical (usually Gaussian) distribution of the deviations from the best fit relation, which were attributed to an intrinsic scatter. Our method, instead, by taking into account all the censored data, allows to measure the {\it shape} of the distribution of the intrinsic scatter and its possible dependences on other variables (such as the luminosity and redshift in our case).
In Figure \ref{Fig_LRLX}  we compare our best fit solution \#5 with the relations derived by other authors such as Brinkmann et al. (2000) in the soft 0.5-2 keV band\footnote{An X-ray spectral index $\alpha$=0.7 was assumed to convert the 0.5-2 keV luminosities into 2-10 keV luminosities.}, Merloni et al. (2003), Panessa et al. (2007), Li et al. (2008) and Ballantyne (2009; for the radio quiet AGN) in the 2-10 keV band. The relations by Merloni et al. (2003) and Li et al. (2008)
were derived from their measure of the AGN fundamental plane (i.e. including also a dependence on the BH mass), assuming a fixed BH mass of 10$^8$ M$_\sun$.

As our measured distribution function is a-symmetrical and depends on the redshift, the comparison is not straightforward (in Figure \ref{Fig_LRLX}  we plot the position of the median R$_0$ of the distribution at redshifts 0 and 3). 
It results that our best fit solution \#5 of the dependence of the 1.4 GHz luminosity on
the 2-10 keV luminosity is flatter than obtained in previous works.
Our fitted relation corresponds to a power-law  L$_R$$\propto$L$_X^\alpha$ with indexes $\alpha$=0.48 and $\alpha$=0.58 at redshifts 0 and 3, respectively. The relations fitted by the other authors have instead a wide range of power-law indexes  (0.6$\lesssim\alpha\lesssim1.5$), systematically steeper than our result.
This difference is partly caused by our new method used to measure the L$_R$-L$_X$ relation but it is also caused by our introduction (and measure) of a dependence on the redshift of the average R$_X$ values. In fact, as shown in Figure \ref{Fig_LRLX}, our solution \#4, obtained without
the inclusion of a redshift dependence, results in a steeper slope, having a power-law index  $\alpha=0.75$. Figure \ref{Fig_LRLX} also shows that the fundamental plane by Merloni et al. (2003) has a power-law index ($\alpha$=0.60) close to our best fit estimate ($\alpha$=0.48-0.58), and for BH masses of 10$^8$ M$_\sun$ is located between our best fit relations at redshift 0 and 3.
In the fundamental plane estimate by Merloni et al. (2003)  R$_X$ has also a dependence on the BH mass of the type
$\propto$$0.78\rm{log}(\rm{M}_{BH})$. This implies
that our estimates at redshift 0 and 3 are roughly similar to the fundamental plane measures for BH masses of 10$^7$ and 10$^9$ M$_\sun$, respectively. Therefore, as in our (and all flux limited) samples, high redshift AGN are on average more luminous, and thus probably host on average more massive BH, we can infer that our measure of the increase with the redshift of the median of the R$_X$ distribution is qualitatively in agreement with the AGN fundamental  plan measures where an increase of the average R$_X$ with the BH masses is observed.

%% FIG 09
  \begin{figure}[]
 \centering
  \includegraphics[width=8.3cm, angle=0]{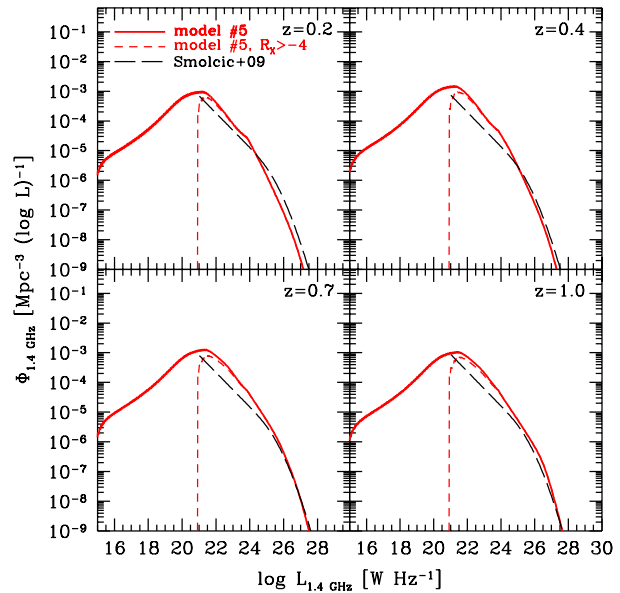}
  \caption{Predicted 1.4 GHz radio LF according to our
  fit \#5, at redshifts 0.2, 0.4, 0.7 and 1.0 (red continuous lines). 
  The dashed red lines show the part of the LF reproduced by the P(R$_X$) distribution with R$_X$$>$-4. The
  black long dashed lines show the radio luminosity function as estimated by Smol\v{c}i\`{c} et al. (2009).}
        \label{Fig_LF5}
 \end{figure}

\subsubsection{The 1.4 GHz luminosity function and counts}

We then verified if the measured probability distribution function of R$_X$, once convolved with the 2-10
keV LF,  properly reproduces (taking into account the uncertainties)  independent previous measures of the
1.4 GHz LF, $\Phi_R(L_{1.4},z)$, and integral counts, $N(>S)$. We used the 2-10 keV LF, 
$\Phi_X(L_X, z)$, as measured by La Franca et al. (2005), and modified by
allowing a steep exponential decline of the AGN density at redshifts larger than $z=2.7$, as measured by Brusa et al. (2009a).  As discussed in La Franca et al. (2005), the density of Compton thick AGN 
with 24$<$logN$_H$$\leq$26 cm$^{-2}$
was assumed to be equal to the density of the Compton thin AGN with 22$<$logN$_H$$\leq$24 cm$^{-2}$ (this assumption
resulted to properly reproduce the cosmic X-ray background).
The X-ray LF can be converted  into a radio LF  by the formula:
\begin{equation}
\Phi_R(L_{1.4} ,z) = \int P(R_X | L_X, z)\Phi_X(L_X, z)d{\rm log}L_X.
\label{eq_rlf}
\end{equation}
The X-ray LF was integrated starting from an X-ray luminosity logL$_X$=41 erg/s.
The predicted radio LF is shown in Figure \ref{Fig_LF5}  and compared with the 1.4 GHz
radio LF measured by Smol\v{c}i\'{c} et al. (2009).  As discussed in Smol\v{c}i\'{c} et al. (2009), this radio
LF includes (like in our measure of the distribution of R$_X$) mostly the FRI sources. 
Although also affected by the uncertainties on the measure of the
X-ray LF by La Franca et al. (2005), our best fit \#5 of the probability
distribution function of R$_X$ provides a fairly good reproduction of the FRI radio luminosity function.

As a result of this computation, our estimate of the distribution of R$_X$ allows to predict the AGN FRI radio LF at luminosities smaller than L$_{1.4}$=10$^{21}$ W Hz$^{-1}$, never probed before. 
However at these low radio luminosities the AGN could be overwhelmed by the radio emission of  the hosting galaxy because during periods of strong star formation activity the supernovae remnants accelerate
cosmic rays which radiate synchrotron emission in local magnetic fields.
According to Ranalli et al. (2003), 
strong star forming galaxies ($\sim$10$^2$ M$_\sun$/yr) have about  10$^{42}$ erg/s 2-10 keV luminosities, and, in general, 
in all star forming galaxies a linear relation between the 1.4 GHz radio and
the 2-10 keV  luminosites is observed, which corresponds to a value R$_X$$\simeq-$2.0.
At face value this would imply that for the lowest luminosity AGN (logL$_X$$\simeq$42-43 erg/s) most of our measure of the P(R$_X$) distribution (which spans in the range $-$7$<$R$_X$$<-$1) could be contaminated  if the hosting galaxies are undergoing a strong
star formation activity. However, as shown in Figure \ref{Fig_distr5}, at low R$_X$ values
 the P(R$_X$) distribution  is mostly measured from AGN with 43$<$logL$_X$$<$45 erg/s,
 and redshift lower than $z$$\sim$0.5. At these low redshifts it is very unlikely for galaxies to harbor 
a star formation stronger than $\sim$10 M$_\sun$/yr (Elbaz et al. 2007; Noeske et al. 2007), which roughly corresponds to an emission in the  2-10 keV band of logL$_X$$\sim$41 erg/s (Ranalli et al. 2003). Therefore,
for example, in a low redshift AGN with logL$_X$$\sim$44 erg/s, having an
hosting galaxy whose star formation emits a 2-10 keV luminosity of logL$_X$$\sim$41 erg/s, the
corresponding radio emission (having an intrinsic R$_X$$\simeq$-2.0) can contaminate our measure of the R$_X$ distribution only for  R$_X$ values smaller than -5. We can then conclude that only at the lowest R$_X$ values ($\lesssim-5$) our measure of the P(R$_X$) distribution is potentially affected by a  contamination from the radio emission due to the star formation activity of the AGN hosting galaxies.

%% Fig 10
   \begin{figure}
 \centering
  \includegraphics[width=6.5cm, angle=-90]{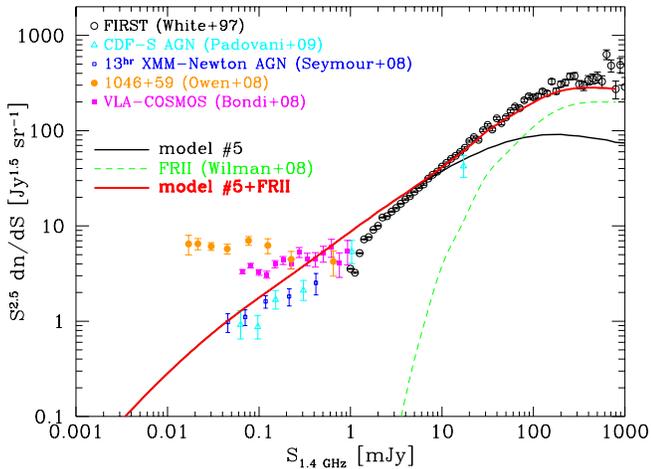}
  \caption{Euclidean  differential radio counts at 1.4 GHz. The black continuous line shows the expected counts according to
  the convolution of the P(R$_X$) distribution  of fit \# 5 (see Table \ref{tab_fit}) with the X-ray LF of La Franca et al. (2005).
  The green dashed line shows the radio counts produced by the FRII sources according to Wilman et al. (2008).
  The red continuous line is the sum of our predicted counts (the black line) with the counts of the FRII sources
  (green dashed line). Radio counts measures are shown by black open circles (FIRST; White et al. 1997), magenta filled squares (Bondi et al. 2008)
  and orange filled circles (Owen et al. 2008). Blue open squares  and cyan open triangles show the AGN counts estimated by Seymour et al. (2008) and Padovani et al. (2009), respectively. \medskip}
        \label{Fig_Counts5}
 \end{figure}

We used the above computed radio LF (eq. \ref{eq_rlf}) to derive the expected integral counts N($>$S) from the following equation:
\begin{equation}
N(>{\rm S})={1\over{4\pi}} \int {{dv}\over{dz}}dz  \int^{L_{max}} _{Sk(z)4\pi d^2_l(z)}
\Phi(L_{1.4}, z)d{\rm log}L_{1.4},
\end{equation}
where $k(z)=(1+z)^{\alpha_R-1}$ (with $F_\nu \propto\nu^{-\alpha_R}$  and $\alpha_R=0.5$)  is the radio k-correction, $d_l(z)$ the luminosity distance, and the integral counts are measured in sr$^{-1}$ units.
The predicted radio counts are shown in Figure \ref{Fig_Counts5}. As our measured distribution of R$_X$  represents only the FRI population (see discussion in sect. 2), in order to reproduce the total radio counts, the contribution derived from the LF of the FRII population (as measured by Wilman et al. 2008) was added.
The reproduced counts are in good agreement with the observations.  At 1.4 GHz fluxes below 1 mJy
the euclidean radio counts flatten, due to the appearance of the  population of the star forming galaxies.
In this context, it is matter of discussion which is the fraction of the AGN at these fluxes.
Our results agree with recent estimates of the AGN contributions to the sub-milliJansky radio counts
from Seymour et al. (2008) and Padovani et al. (2009).  Similar results (at these fluxes) have also been obtained from semi-empirical simulations of the extragalactic radio counts by Jarvis \& Rawlings (2004),
Wilman et al. (2008) and Ballantyne (2009).

\section{The kinetic luminosity function}
%% FIG 11
  \begin{figure}[]
 \centering
  \includegraphics[width=6.1cm, angle=-90]{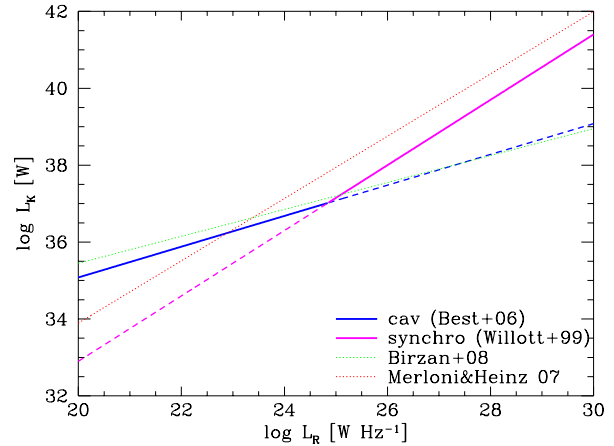}
  \caption{Kinetic luminosity as function of the radio luminosity as estimated by
  Willott et al. (1999; magenta lines), Best et al. (2006; blue lines), Merloni \& Heinz (2007; red dotted line) and
  Birzan et al. (2008; green dotted line). In our work we used the combination of the relations from Best et al. (2006) and
 Willott et al. (1999) at low and high luminosities, respectively (the two continuous lines).
\medskip }
        \label{Fig_LRLK}
 \end{figure}

Once the radio LF of the FRI sources is measured (Equation \ref{eq_rlf} and Figure \ref{Fig_LF5}), in order to derive the kinetic mechanical LF and its evolution, we should convolve the 1.4 GHz LF with  a relation which converts the radio luminosity L$_{1.4}$ into a mechanical power L$_K$.
In the last decade, several authors have worked on the estimate of this relation (Willott
 et al. 1999; Birzan et al. 2004, 2008; Best et al. 2006; Heinz et al.  2007; Merloni \& Heinz 2008). Following the discussion of Cattaneo and Best (2009) we
 used two different estimates which are representative of two different luminosity regimes. At high radio luminosities (L$_{1.4}$$\ga$10$^{25}$ W Hz$^{-1}$), Willott et al. (1999) used the minimum entropy density that the plasma radio lobes must have in order to emit the observed synchrotron radiation and obtained (see Figure \ref{Fig_LRLK}):
  \begin{equation}
L_{\rm K} = 1.4 \times 10^{37} \left(\frac{L_{\rm
  1.4\,GHz}}{10^{25}{\rm W\,Hz}^{-1}}\right)^{0.85}{\rm W}.
\label{w2}
\end{equation}

A second approach is to infer $L_{\rm K}$ from the mechanical work that
the lobes do on the surrounding hot gas. The expanding lobes of
relativistic synchrotron-emitting plasma open cavities (of volume $V$) in the ambient
thermal X-ray emitting plasma. The minimum work in
inflating these cavities is done for reversible (quasi-static) inflation
and equals $pV$, where $p$ is the pressure of the ambient gas.
Best et al. (2006) derived a relation between radio and mechanical
luminosity based upon this estimate for the energy associated with these
cavities, combined with an estimate of the cavity ages from the buoyancy
timescale (from Birzan et al. 2004).  Comparing the mechanical
luminosities of 19 nearby radio sources that have associated X-ray cavities
with their 1.4\,GHz monochromatic radio luminosities leads to a relation
 \begin{equation}
L_{\rm K} = 1.2 \times 10^{37} 
\left(\frac{L_{\rm 1.4\,GHz}}{10^{25}{\rm W\,Hz}^{-1}}\right)^{0.40}
{\rm W},
\label{b2}
\end{equation}
 which is better suited for low luminosities (L$_{1.4}$$\la$
10$^{25}$ W Hz$^{-1}$) and is close to the estimate by Birzan et al. (2008; see Figure \ref{Fig_LRLK}).
Using a similar method, Merloni \& Heinz (2008) obtained at high radio luminosities, a relation similar to the
one by Willott et al. (1999) but with about 0.5-1 dex higher kinetic luminosities (see Figure \ref{Fig_LRLK}).

\begin{figure}[]
 \centering
  \includegraphics[width=8.5cm, angle=0]{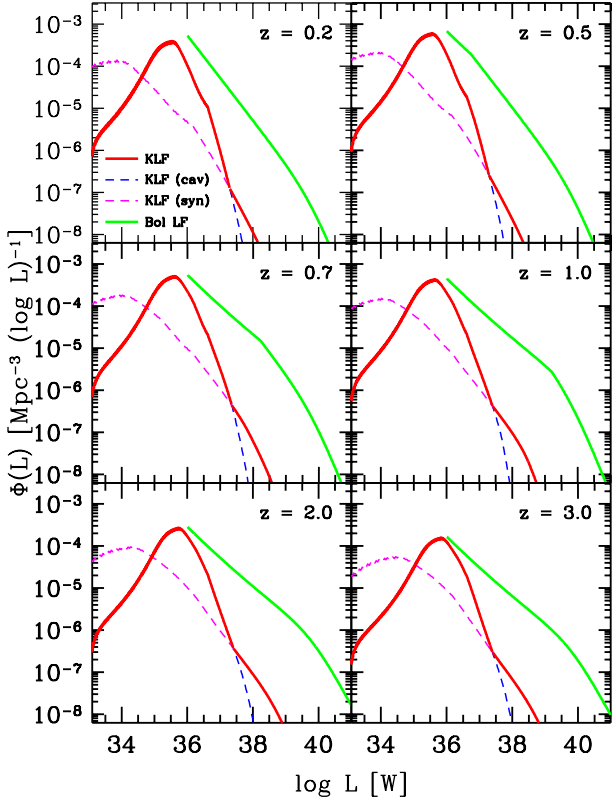}
  \caption{Bolometric  radiative (green continuous line) and kinetic AGN LF. The magenta and blue lines show the kinetic LF predicted using the relations linking the radio and kinetic luminosities  of
Willott et al. (1999) and Best et al. (2006) respectively. The red line is the result obtained using the combination of the relations of Willott et al. (1999) and Best et al. (2006) at high and low luminosities respectively (see text and Figure \ref{Fig_LRLK}).
  }
        \label{Fig_KLF}
 \end{figure}

\begin{figure}[]
 \centering
  \includegraphics[width=8.5cm, angle=0]{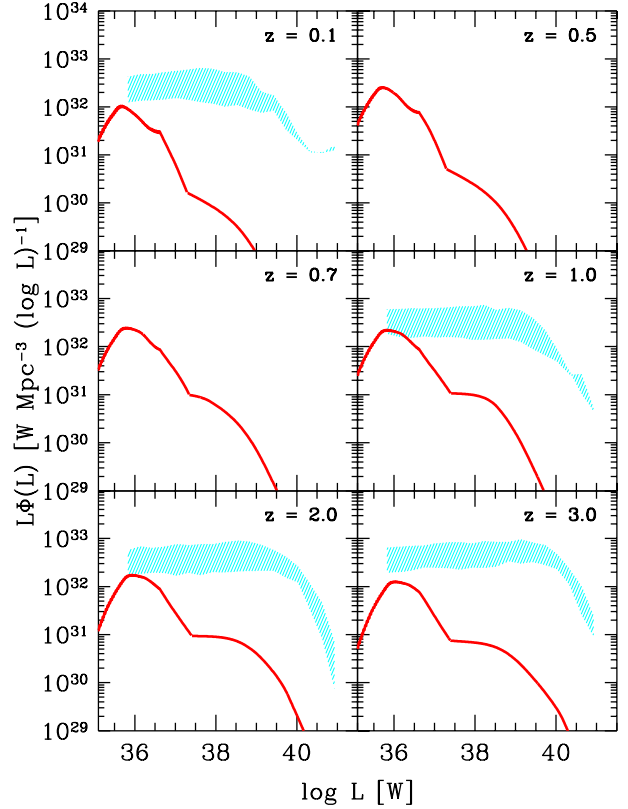}
  \caption{Kinetic power density, L$_K$$\Phi_K$(L$_K$,$z$), as function of the kinetic luminosity according to the fit solution \#5 (red continuos line). 
  The results by Merloni \& Heinz (2008) are shown by a cyan shaded area.}
        \label{Fig_LKLF}
 \end{figure}

In this work (as in Cattaneo and Best 2009) we used Equation \ref{w2} at high luminosities (L$_{1.4}$$\ga$10$^{25}$ W Hz$^{-1}$) and Equation \ref{b2} at lower luminosities.
Using these relations, once derived the radio LF from the X-ray LF and the R$_X$ distribution according to Equation 4, we can estimate the kinetic LF, $\Phi_K(L_K, z)$, by the formula
\begin{equation}
\Phi_K(L_K, z)= { d N(L_K, z) \over{ dV d{\rm log} L_K}}= \Phi_{1.4}(L_{1.4}(L_K), z) { d {\rm log}L_{1.4} \over{ d{\rm log} L_K}},
\end{equation}
while the bolometric radiative LF, $\Phi_{rad}(L_B, z)$, is derived from the X-ray LF, via the bolometric correction $L_B = K_X(L_X)L_X$, by 
\begin{equation}
\Phi_{rad}(L_B, z)= { d N(L_B, z) \over{ dV d{\rm log} L_B}}= \Phi_X(L_X(L_B), z) { d{\rm llog} L_{X} \over{ d{\rm log} L_B}},
\end{equation}
where for K$_X$(L$_X$) we used the relation from  Marconi et al. (2004). 
Both the, so derived, kinetic and radiative (bolometric) LF are shown in Figure  \ref{Fig_KLF}, while in Figure
 \ref{Fig_LKLF} the kinetic power density, L$_K$$\Phi_K$(L$_K$,$z$), as function of the kinetic luminosity is shown.

 %%%% FIG 12
 
%%%% FIG 13

\begin{figure}[]
 \centering
  \includegraphics[width=8.5cm, angle=0]{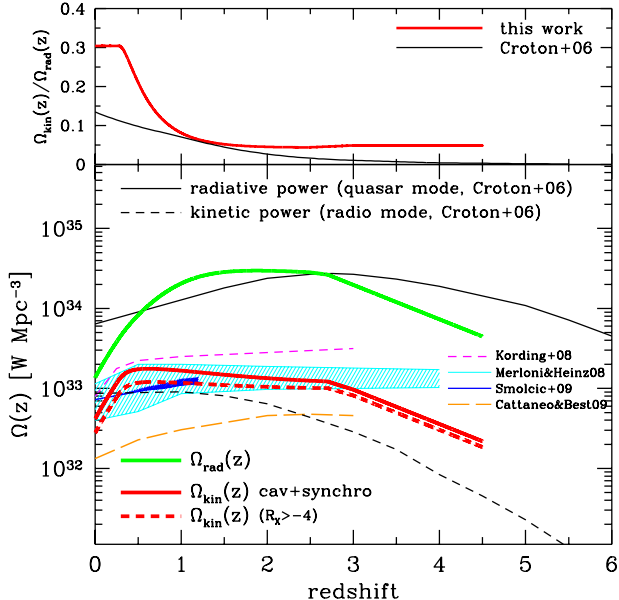}
  \caption{{\it Bottom}. Power density as a function of redshift.  The
    kinetic power density derived from the best fit solution \#5 is
    shown by a continuous red line.  The red dashed line shows the
    result if the more radio loud objects (R$_X$$>$-4) only are used
    (see text and Figure \ref{Fig_LF5}).  The magenta and orange
    dashed lines show the kinetic luminosity density as estimated by
    Kording et al. (2008) and Cattaneo \& Best (2009),
    respectively. The cyan and blue areas show the estimates by
    Merloni \& Heinz (2008) and Smol\v{c}i\'c et al. (2010),
    respectively.  The radiative and kinetic power densities, as used
    in the model of Croton et al. (2006), are shown by continuous and
    dashed black lines, respectively. {\it Top}.  Ratio between the
    kinetic and radiative power densities. The red and black
    continuous lines are the results by our fit \#5 and Croton et
    al. (2006), respectively.}
        \label{Fig_Kin1}
 \end{figure}

 By integrating in luminosity  
the above derived kinetic and radiative (bolometric) LF it is possible to estimate the dependence of the AGN mechanical and radiative power per unit cosmic volume as a function of redshift,

\begin{equation}
\Omega_K(z)= \int L_K(L_{1.4}) \Phi_{1.4}(L_{1.4},z) d{\rm log}L_{1.4},
\label{Eq_Ok}
\end{equation}

and
\begin{equation}
\Omega_{rad}(z)= \int L_{bol}(L_X)\Phi_X(L_x,z)d{\rm log}L_X,
\label{Eq_Or}
\end{equation}
which are shown in Figure \ref{Fig_Kin1}. As both the above equations depends on the X-ray luminosity function
(see eq. \ref{eq_rlf})  the integral lower limit was logL$_X$=41 erg/s which corresponds to a lower limit of 
of logL$_B$$\simeq$35.5 W for eq. \ref{Eq_Or} and, according to the R$_X$ distribution and the relation between L$_{1.4}$ and L$_K$ (eq. \ref{b2}) to a lower limit of 
logL$_K$$\simeq$34.5 W for eq. \ref{Eq_Ok}.
The resulting kinetic power density would be similar if, at all luminosities, the conversion of the radio into kinetic luminosity by Best et al. (2006),  eq. \ref{b2},  would be used, while a factor 4-6 lower values would be obtained using the relation from Willott et al. (1999),  eq. \ref{w2}, only.  As shown in Figure \ref{Fig_KLF}
 this difference is caused by the steep drop off of the 1.4 GHz radio LF at luminosities higher than LogL$_R$$\sim$25 W (see Figure \ref{Fig_LF5}); as a consequence the resulting power density (the integrated kinetic LF) depends mainly on which  conversion of the radio luminosity into kinetic power
 is used at low luminosities.

%% Fig 14

\section{Discussion}

%%%%%%%%%%%%%%%%%%%%%%%%%%%%%%%%%%%%%%%%%%%%%%%

The probability distribution function of R$_X$, $P(R_X|
L_X, z)$ estimated in Section 3 depends on both luminosity and redshift: the
average R$_X$ increases with decreasing luminosity and (possibly) increasing
redshift (see best fit \#5 in Table \ref{tab_fit} and Figure
\ref{Fig_LZ5}).  The observed increase of the average R$_X$ value with
decreasing luminosity is similar to previous results and models where
(in analogy with X-ray binaries) low luminosity AGN are expected to be
more likely radio loud 
\footnote{As discussed in sect. 2 and 3, these trends do not include the most luminous (radio loud; mostly FRII) sources.
As far as the more radio loud population is concerned,
it has been observed that the fraction of more radio loud AGN
increases with increasing optical (or X-ray) luminosity and decreasing redshift
(Miller et al. 1990; Visnovsky et al. 1992; Padovani 1993;
La Franca et al. 1994, Goldschmidt et al. 1999, Ivezi{\'c} et al. 2002;
Cirasuolo et al. 2003; Jiang et al. 2007).} (see e.g. Merloni \& Heinz 2008; K\"ording,
Jester \& Fender 2008).

The knowledge of the R$_X$ distribution, once convolved with the X-ray
LF and the relations between the kinetic and radio luminosity, allows 
to estimate the kinetic LF and its evolution.  At luminosities higher
than the break of the bolometric LF (L$_K$$\sim$10$^{39}$ W) the
kinetic LF results to be more than two orders of magnitude smaller than
the bolometric LF (see Figure \ref{Fig_KLF}), while at lower
luminosities the relevance of the kinetic LF increases, reaching values
comparable to the bolometric one at L$_K$$\sim$10$^{36}$ W, which
roughly corresponds to the minimum luminosity experimentally probed by
the X-ray LF (logL$_X$$\sim$42 erg/s; see e.g. La Franca et al. 2005).
The kinetic LF shows a maximum in the range
10$^{35}$$<$$L_{K}$$<$10$^{37}$ W, where most ($\sim$90\%) of the
kinetic power density (shown in Figure \ref{Fig_LKLF}) is produced.

In Figure \ref{Fig_Kin1} we show the kinetic and radiative power
density as a function of the redshift.  We also show the kinetic power
density corresponding to the more radio emitting AGN having $R_X$
larger than -4, and then corresponding to the population which is
typically represented by the radio LF (see in Figure \ref{Fig_LF5} the
comparison of the reproduced radio LF with R$_X$$>$-4 with the radio
LF of FRI sources as measured by Smol\v{c}i\'c et al. 2009).  It is
then possible to see that the kinetic power density could be
underestimated by up to a factor of about two, if the radio LF alone is
used, without taking into account the low radio luminosity AGN population.

Our estimates are in qualitative agreement with the trends of the
radiative and kinetic power density with redshift used by Croton et al
(2006) at $z>0.5$ (see Figure \ref{Fig_Kin1}). However, at lower
redshifts we find a sharp (a factor of five) decrease of both radiative and kinetic power
densities  from $z\sim$0.5 (i.e. about 5 billion years ago) to $z=0$. This result is
quite robust, and comes from the strong negative evolution of the AGN
LF from $z\sim2$ down to $z=0$, which has been observed since the
first studies of the QSO evolution in the optical (see e.g. Marshall
et al. 1983; Croom et al. 2009 for recent results), and measured by
many other authors in the hard X-rays (e.g. Ueda et al. 2003; La
Franca et al. 2005; Hasinger 2008). Conversely, Croton et al. (2006)
assume an almost continuous increase of both the kinetic and radiative powers,
due to the assumption that both phenomena are related to an almost constant
accretion onto the SMBH. Under this assumption Croton et al. (2006)
overestimate both the AGN radiative and
kinetic power densities at low ($z$$\lesssim$0.5) redshift,  allowing only
for a much shallower decrease 
of the kinetic feedback (a factor of 30\%)  and of  the 
AGN radiative power.

In Figure \ref{Fig_Kin1} ({\it top}) we show the ratio of the
$\Omega_K/\Omega_{rad}$ plotted as function of redshift. According to
our best fit \#5, at $z$$>$0.5 the kinetic power density is $\sim5\%$ of
the radiative density, and increases up to about 30\% at decreasing
redshifts.  This increase at low redshifts of the ratio of the kinetic
power to the radiative power density could help in modeling the
quenching of the star formation at low redshift. In Croton et
al. (2006) a milder increase is assumed, which should be attributed to
the, above discussed, overestimate of the low redshift AGN radiative
density.

Many previous results on the AGN kinetic LF are based on
the convolution of some relations between the kinetic and radio power
with direct measures of the AGN radio LF (e.g. Shankar et al. 2008; Merloni \& Heinz 2008;
K\"ording, Jester \& Fender 2008; Cattaneo \& Best 2009; Smol\v{c}i\'c
et al. 2009). As already discussed in the introduction, the measure of
the R$_X$ distribution is useful in order to allow a detailed
implementation of the AGN feedback within the galaxy formation and
evolution models because it gives the opportunity to predict the radio
luminosity (and thus feedback) of each AGN as a function of its
luminosity (accretion rate) and redshift.

According to our best fit \#5, in the redshift range
0.5$<$$z$$<$3 the integrated kinetic power density is
$\sim$1-2$\times 10^{33}$ W Mpc$^{-3}$ (see Figure \ref{Fig_Kin1}). This is in rough agreement
with the previous estimates by Merloni \& Heinz (2008), K\"ording,
Jester and Fender (2008) and Smol\v{c}i\'c et al. (2009). At lower
redshift ($z$$<$0.5) we observe a drop by a factor of five, similar to
what observed by K\"ording, Jester and Fender (2008), and in
agreement, within the uncertainties, with Merloni \& Heinz (2008). On
the contrary our results are, at any redshift, 2-8 times greater than
that reported by Cattaneo \& Best (2009).

Merloni \& Heinz (2008) found that their kinetic LF roughly
corresponds to a constant overall efficiency in converting the
accreted mass energy into kinetic power $\epsilon_{kin}$$\simeq$$ 3-5
\times10^{-3}$ (where $L_K=\epsilon_{kin}\dot{m}c2$).  Their results
are similar to ours where, according to equations \ref{Eq_Ok} and
\ref{Eq_Or}, on average, we have $\epsilon_{kin}$$\simeq$$
(\Omega_K/\Omega_R) \epsilon_R\simeq 5\times 10^{-3}$, as we measure
$\Omega_K/\Omega_R\simeq 0.05$ (see the $\Omega_K/\Omega_R$ ratio as a
function of redshift in Figure \ref{Fig_Kin1}, {\it top}), and
assuming a radiative efficiency $\epsilon_R=0.1$ (Marconi et
al. 2004). However, we observe an increase of the $\Omega_K/\Omega_R$
ratio (i.e. of the kinetic efficiency) up to a value 0.3 at decreasing
redshifts, which Merloni \& Heinz (2008) observe in the most massive
objects only ($> 10^{8}-10^{9} M_\sun$; see e.g. their Figure 13).  It
should also be noted that, although the integrated kinetic power
density of Merloni \& Heinz (2008) is in agreement with our estimate,
their kinetic LF is similar to our measure only at
L$_K$$\sim$10$^{36}$ W (see Figure \ref{Fig_LKLF}), while it is
definitely larger (by about an order of magnitude) at higher kinetic
luminosities. Once integrated in luminosity, the computed power
densities are similar (at $z$$\lesssim$3) because our low luminosity
limit (L$_K$=10$^{34}$ W) is significantly lower than that used by
Merloni \& Heinz (2008; L$_K$=10$^{36}$ W).

Shankar et al. (2008), found that the ratio of the kinetic to
bolometric luminosity, defined as $g_k=L_K/L_B$, is constant and equal to
$g_k=0.10$ with a scatter of $\sigma=0.38$.
According to equations \ref{Eq_Ok} and \ref{Eq_Or}, $g_k$ corresponds
roughly to the ratio $\Omega_K/\Omega_{rad}$ (plotted as function of
redshift in Figure \ref{Fig_Kin1}, {\it top}), which, as discussed
above, levels at $\sim$0.05 at $z>1$ while increases up to 0.3 at $z=0$.

\section{Conclusions}

We used a sample of more than  1600 X-ray selected AGN observed at 1.4 GHz to measure the probability distribution function, $P(R_X| L_X, z)$, of the ratio $R_X$ of the radio to intrinsic X-ray luminosity, as a function of the X-ray luminosity and redshift. 

The  knowledge of the $P(R_X| L_X, z)$ distribution is necessary to estimate the AGN kinetic (radio) feedback into the hosting galaxies
by allowing to couple it with the luminous, accreting, phases of the AGN activity.

The average value of R$_X$ increases with decreasing X-ray luminosities and (possibly) increasing redshift. At variance, we did not find a statistical significant difference between the radio properties of the X-ray absorbed (N$_H$$>$$10^{22}$ cm$^{-2}$) and un-absorbed AGN.

We were able to better measure the densities of the more radio quiet  (R$_X$$<$-4) AGN which resulted to be responsible of
about half of the derived kinetic power density.

According to our analysis the value of the kinetic energy density is in qualitative agreement with the last generation galaxy evolution  scenarios, where radio mode AGN feedback is invoked to quench the star formation in galaxies and slow down the cooling flows in galaxy clusters.
However at redshifts below 0.5,
similarly to what observed by K\"ording, Jester and Fender (2008),
 we find a sharp (about a factor of five) decrease of the kinetic energy density,
 which is strictly related the AGN density evolution, but  which is not included in many of the galaxy/AGN formation and
 evolution models where, instead, the radio mode feedback  is assumed to continuously increase (or only smoothly decrease) at low redshift.

\acknowledgments

We thank A. Lamastra, N. Menci, R. Morganti, P. Ranalli, V. Smol{\v c}i{\'c} and G. Zamorani for discussions.
We are grateful to L. Trouille and A. Barger for the support in allowing us to use the CLANS data,  and to D. Ballantyne, R. Della Ceca and A. Merloni for providing data in machine readable format. We thanks the anonymous referee
for his useful comments. 
We acknowledge financial contribution from contract ASI-INAF I/088/06/0 and PRIN-MIUR grant 2006-02-5203.


%\noindent
%{\it Facilities:} \facility{VLT:Melipal (VIMOS)}, 
%\facility{VLT:Antu (FORS2)}, \facility{ESO:3.6m ()},
%\facility{Spitzer ()}%%, \facility{XMM ()}

\vfill\eject

\end{document}